\documentclass[a4paper,11pt]{article}
\usepackage{pos}
\usepackage{hyperref}
\usepackage{url}
\usepackage{amsmath}
\usepackage{graphicx}
\usepackage{multirow}
\usepackage{microtype}

\newcommand\beq{\begin{eqnarray}}
\newcommand\eeq{\end{eqnarray}}
\newcommand\Tab[1]{Table~\ref{tab:#1}}
\newcommand\Fig[1]{Fig.~\ref{fig:#1}}

\title{Vector-channel scattering of dark particles in a Sp(4) gauge theory}
\author*[a]{Jong-Wan Lee}
\author[b,c]{Ed Bennett}
\author[d]{Yannick Dengler}
\author[e,f]{Deog Ki Hong}
\author[g]{Ho Hsiao}
\author[h,i]{C.-J. David Lin}
\author[b,j]{Biagio Lucini}
\author[d]{Axel Maas}
\author[c,k]{Maurizio Piai}
\author[l]{Davide Vadacchino}
\author[c,k,m]{Fabian Zierler}

\affiliation[a]{Center for Theoretical Physics of the Universe,
 Institute for Basic Science, Daejeon 34126, Korea }

\affiliation[b]{Swansea Academy of Advanced Computing,
Swansea University (Bay Campus), Fabian Way, Swansea SA1 8EN, United Kingdom }

\affiliation[c]{Centre for Quantum Fields and Gravity, Faculty  of Science and Engineering, Swansea University, Singleton Park, SA2 8PP, Swansea, United Kingdom}

\affiliation[d]{Institute of Physics, 
NAWI Graz, University of Graz, Universitätsplatz 5, A-8010 Graz, Austria }

\affiliation[e]{Department of Physics, 
Pusan National University, Busan 46241, Korea }

\affiliation[f]{Extreme Physics Institute, Pusan National University, Busan 46241, Korea}

\affiliation[g]{Center for Computational Sciences, 
University of Tsukuba, 1-1-1 Tennodai, Tsukuba, Ibaraki 305-8577, Japan }

\affiliation[h]{Institute of Physics, 
National Yang Ming Chiao Tung University, 1001 Ta-Hsueh Road, Hsinchu 30010, Taiwan }

\affiliation[i]{Centre for High Energy Physics, Chung-Yuan Christian University, Chung-Li 32023, Taiwan}

\affiliation[j]{School of Mathematical Sciences, 
Queen Mary University of London, Mile End Road, London, E1 4NS, United Kingdom }

\affiliation[k]{Department of Physics, 
Faculty of Science and Engineering, Swansea University, Singleton Park, SA2 8PP, Swansea, United Kingdom}

\affiliation[l]{Centre for Mathematical Sciences, 
University of Plymouth, Plymouth, PL4 8AA, United Kingdom }

\affiliation[m]{Technical University of Munich, TUM School of Natural Sciences, Physics Department, James-Franck-Str. 1, 85748 Garching, Germany}

\emailAdd{j.w.lee@ibs.re.kr}

\abstract{
\begin{center}
\href{https://telos-collaboration.github.io}{ \includegraphics[height=1cm]{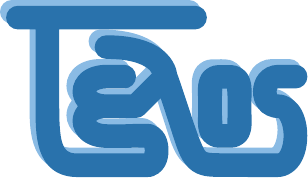}}\\
(on behalf of the TELOS collaboration)
\end{center}

We report new results obtained in our lattice studies of the $Sp(4)$ gauge theory coupled to two fundamental Dirac fermions. This theory provides a candidate for the dynamical origin of dark matter models within the strongly interacting massive particle paradigm. We employ Lüscher's formalism to analyse finite-volume energy levels and study the scattering amplitude of two pseudoscalar states in the spin-1 channel. We present our preliminary findings for a set of ensembles generated within a broad range of (Wilson) fermion masses.
}

\FullConference{
The 42nd International Symposium on Lattice Field Theory (LATTICE2025),  
2-8 November 2025\\
Tata Institute of Fundamental Research, Mumbai, India
}

\begin{document}
\begin{flushright}
\textsc{CTPU-PTC-26-08, TUM-EFT 206/26}
\vspace{-1.5cm}
\end{flushright}

\maketitle

\section{Introduction}

Astrophysical evidence, from galactic rotational curves, bullet cluster, and gravitational lensing, indicates the existence of dark matter in our universe, yet its nature is still unknown, despite numerous candidates with theoretical explanations proposed in the literature~\cite{Cirelli:2024ssz}. Direct and indirect experimental searches for dark particles put stringent bounds on electroweak-scale models, such as those expected in the weakly interacting massive particles (WIMPs) scenario.  A compelling alternative was proposed in Ref.~\cite{Hochberg:2014dra},  by postulating the existence of a dark sector comprised of strongly interacting massive particles (SIMPs), with mass at a sub-GeV scale. Such a scenario makes distinctive prediction for small-scale structure formation, addressing such puzzles  as the spherical properties of dark matter cores, the core-vs-cusp problem, and the too-big-to-fail question~\cite{Tulin:2017ara}. 

In the SIMP scenario, dark matter is identified with the pseudo-Nambu-Goldstone-Bosons (pNGBs) arising from the dynamical breaking of a global continuous symmetry in the dark sector. The distinctive feature of SIMPs is that the thermal relic density is predominantly obtained by a number-changing process in the dark sector, in the form of $3\to 2$ annihilations, rather than the freeze-out mechanism in which two dark matter particles annihilate into two standard-model ones. 

As discussed in Ref.~\cite{Hochberg:2014kqa}, the $Sp(2N)$ gauge theory with $N_f=2$ Dirac fundamental flavours stands out as a minimal model compatible with astronomical observations and theoretical constraints. The number changing process arises from the $5$-point pNGBs interactions associated with the Wess-Zumino-Witten (WZW) anomaly term.  The self-interactions between pNGBs control $2\to2$ scattering, addressing the small-scale problems. The maximum value of self-scattering cross section is constrained by, for instance, observations of the Bullet-cluster, yielding  
$\sigma_{\rm scatter}/m_{\rm DM} \lesssim 1{\rm cm^2}/g$ with $m_{\rm DM}$ the dark matter mass~\cite{Tulin:2017ara}, which can be accommodated by relatively large pNGB masses. 

The combination of these considerations indicates that in the interesting region of parameter space one needs to take into account effects that go beyond the leading-order effective field theory treatment. Heavy resonances, in particular the vectors, have an important role to play, as they appear in a channel shared by both $2\to2$ scattering and $3\to2$ semi-annihilation. Their treatment with hidden local symmetry can be used to understand the bullet-cluster bounds~\cite{Choi:2018iit}  (see also Ref.~\cite{Kulkarni:2022bvh}). 

We performed numerical studies of $Sp(4)$ lattice theory coupled to two (mass-degenerate)  Wilson-Dirac fermions transforming in the fundamental representation of the gauge group, focusing on the scattering of two pNGBs in the spin-1 channel.\footnote{See also Ref.~\cite{Bennett:2019jzz}, for earlier lattice studies of  meson spectra in the same theory, and the use of leading-order approximation of the Wilson chiral perturbation theory for  continuum and massless extrapolations.} We construct the interpolating operators in the spin-$1$, parity-odd channel combining one and two fermion-bilinear operators, with non-vanishing  (small) momenta. We measure the low-lying energy eigenvalues by performing a variational analysis. We then apply Lüscher's finite volume analysis in the elastic scattering regime and extract the corresponding $p$-wave phase shift in the infinite volume theory~\cite{Luscher:1986pf,Luscher:1991cf}.  Focusing on three choices of lattice parameters in the region of parameter space that is relevant to phenomenological studies of SIMP dark matter, we analyse the scattering data using two complementary low-energy approximations, the effective range expansion and the Breit-Wigner form for a resonance. 

The results reported here are based on a preliminary analysis. Further analysis and data generated for this manuscript will be released together with an upcoming publication \cite{TELOS:2026}.\footnote{Alternatively, in the interim, preliminary data and code can be obtained from the authors upon request.}

\section{The model}

The Lagrangian density of the $Sp(4)$ theory coupled to two fundamental Dirac fermions is 
\beq
{\cal L}=-\frac{1}{2}{\rm Tr}\, F_{\mu\nu} F^{\mu\nu} 
+\overline{Q^i}(i\gamma^\mu D_\mu-m) Q^i,
\eeq
where the summation over the flavour index, $i=1,\,2$, is understood, and we omit spinor and colour indexes.
The field-strength tensor and covariant derivative, $F_{\mu\nu}=\partial_\mu A_\nu -\partial_\nu A_\mu + ig \left[A_\mu,A_\nu\right]$ 
and $D_\mu Q = \partial_\mu Q + ig A_\mu Q$, are written in terms of the matrix-valued gauge fields, 
 $A_\mu= A^A_\mu T^A$, where $T^A$ are the generators of $Sp(4)$ satisfying 
${\rm Tr}\, T^A T^B =\frac{1}{2}\delta^{AB}$ with $A,\,B=1,\,\dots, 10$, while $g$ is the gauge coupling.
Because the fundamental representation of $Sp(4)$ is pseudo-real, 
the global symmetry is enhanced from $U(1)\times SU(2)\times SU(2)$ to $SU(4)$---ignoring the anomalous $U(1)_A$. 
The mass term and the fermion condensate, $\Sigma=\langle Q \overline{Q} \rangle$, written in terms of 2-component spinors, are proportional to the $4\times 4$ symplectic matrix, $\Omega\equiv\left(\begin{smallmatrix}
        \mathbf{O}_{2\times 2}  & \mathbf{1}_{2\times 2} \\
        -\mathbf{1}_{2\times 2} & \mathbf{O}_{2\times 2}  
    \end{smallmatrix}\right)$, 
and break the global symmetry as $SU(4)\rightarrow Sp(4)$, both explicitly and spontaneously. 
We expect light pNGBs to emerge, spanning the $SU(4)/Sp(4)$ coset, transforming as the $5$-dimensional representation of the unbroken $Sp(4)\sim SO(5)$ symmetry. 

The scattering channel of two pNGBs decomposes as ${\bf 5}\otimes {\bf 5}={\bf 1}\oplus {\bf 10} \oplus {\bf 14}$ in terms of $Sp(4)$ representations. 
Similarly, the system of three pNGBs, relevant to $3\to2$ dark matter annihilation, 
decomposes as ${\bf 5}\otimes {\bf 5}\otimes {\bf 5}=3\cdot {\bf 5}\oplus {\bf 10}\oplus {\bf 30}\oplus 2\cdot{\bf 35}$. 
The ${\bf 10}$ appears in both two- and three-pNGBs processes, and is the multiplet that includes the vector meson, with important phenomenological implications for dark matter, in particular when the vector meson has mass 
close to the  two- and three-pNGB thresholds. 
Hereafter, we use the notation $\pi_{\rm D}$ and $\rho_{\rm D}~(\rho'_{\rm D})$ for the pNGBs and the lightest (first excited) vector states in the theory, not to be confused with the unrelated QCD bound states. 
We focus on $\pi_{\rm D} \pi_{\rm D}$ scattering in the spin-1 channel and  the role of the $\rho'_{\rm D}$ resonance, leaving the $\rho_{\rm D}$ resonance and $\pi_{\rm D}\pi_{\rm D} \pi_{\rm D}$, $\pi_{\rm D}\rho_{\rm D}$ scattering for future work. Lattice studies of the $14$-dimensional channel in the $\pi_{\rm D}\pi_{\rm D}$ scattering are presented in Ref.~\cite{Dengler:2024maq}.

\section{Numerical strategy}

The continuum rotational symmetry is broken to the octahedral group, ${\cal O}_h$, the discrete symmetry of a cube. 
If the system is boosted with non-vanishing momentum, $\vec{P}$, this discrete group is further reduced to its little subgroup (LG).  We consider four different centre-of-mass lattice momenta, $\vec{P}=(2\pi/L) \vec{d}$ with $\vec{d}=(0,0,0)$, $(0,0,1)$, $(1,1,0)$, and $(1,1,1)$, corresponding to the four LGs denoted as
 ${\cal O}_h$, $C_{4v}$, $C_{2v}$, and $C_{3v}$, respectively. 
Rather than using the continuum quantum numbers, $J^P$, with $J$ the spin and $P$ the parity,
 we classify composite and scattering states according to the irreducible representations of the LGs. We restrict attention to $p$-wave scattering amplitudes, and focus on the subduced representations associated with $J^P=1^{-}$, 
neglecting contributions from  higher partial waves ($J\geq 2$). 

The two types of interpolating operators we use are defined as follows:
\beq
\label{eq:ov}
{\cal O}_{V_\mu}(t,\vec{P}) &\equiv&  \sum_{\vec{x}} \overline{Q^i} \Gamma_\mu Q^j e^{i\vec{P}\cdot \vec{x}}, \\
{\cal O}_{\pi_{\rm D}\pi_{\rm D}}(t,\vec{P}) &\equiv&  \frac{1}{\sqrt{2}}\left(
{\cal O}_{\pi_{\rm D}^+}(t,\vec{p}_1) {\cal O}_{\pi_{\rm D}^-}(t,\vec{p}_2) -
{\cal O}_{\pi_{\rm D}^+}(t,\vec{p}_2) {\cal O}_{\pi_{\rm D}^-}(t,\vec{p_1}) 
\right)\,,
\label{eq:o2ps}
\eeq
where in the single meson operator we consider both gamma-matrix structures, $\Gamma_\mu=\gamma_\mu$ and $\gamma_0 \gamma_\mu$ with $\mu=x,\,y,\,z$, 
which interpolate the same states in continuum. The pNGB operators are defined in the analogous way, as ${\cal O}_{\pi_{\rm D}^{+(-)}}(t,\vec{p}) \equiv \sum_{\vec{x}} \overline{Q^{1(2)}}\gamma_5 Q^{2(1)} e^{i\vec{p}\cdot\vec{x}}$,
and the momentum $\vec{P}=\vec{p}_1+\vec{p}_2$ is the sum of the momenta in the two particle-states. Flavour indices $i\neq j$ are understood. We isolate the desired state (irreducible representation), $R$, using the following projection formula ~\cite{Alexandrou:2017mpi}:
\beq
{\cal O}^R(t,\vec{p}) = \frac{d_R}{N_{\rm LG}} \sum_{\tilde{g}\in {\rm LG}} \chi_R (\tilde{g})\, \tilde{g} {\cal O}(t,\vec{p}),
\label{eq:ovr}
\eeq
where $\tilde{g}$ is  the group element,  $N_{\rm LG}$ the order of the little group, while $d_R$ and $\chi_R(\tilde{g})$ are  the dimension and character of the representation, respectively. 
Note that the group elements act on the momentum in the centre-of-mass frame, as well as the intrinsic polarisation vector. 
We summarise the projected operators used in this work in \Tab{ops}.
For the $\pi_{\rm D}\pi_{\rm D}$ operators, we only consider one
specific assignment of the momenta, $\vec{p}_1=0$ and $\vec{p}_2=\vec{P}$, 
in the $A_1$ irreducible representation, due to considerations of numerical cost.

\begin{table}[t]
\centering
\caption{%
Interpolating operators, ${\cal O}$, and matrix elements, ${\cal M}$, used in our spin-1 scattering channel analysis---see Eqs.~(\ref{eq:ov}), (\ref{eq:o2ps}), and (\ref{eq:ovr})~\cite{Gockeler:2012yj,Alexandrou:2017mpi}---together with
the corresponding little group, LG, the vector, $\vec{d}$, defining the total momentum, $\vec{P}=(2\pi/L) \vec{d}$, 
as well as the irreducible representations, $R$, of the octahedral group.
}
\begin{tabular}{|c|c|c|c|c|c|}
\hline
LG & $\vec{d}$ & $R$ & ${\cal O}^R_{V}$ & ${\cal O}^R_{\pi_{\rm D} \pi_{\rm D}}$ & ${\cal M}_{lm,lm}$ \\ \hline \hline
$O_h$ & (0,0,0) & $T_1$ & ${\cal O}_{V_x}+{\cal O}_{V_y}+{\cal O}_{V_z}$  & - & $w_{00}$ \\ \hline
\multirow{2}{*}{$C_{4v}$} & \multirow{2}{*}{(0,0,1)} & $A_1$ & ${\cal O}_{V_z}$ & ${\cal O}_{\pi_{\rm D} \pi_{\rm D}}$ & $w_{00}+2w_{20}$\\
                          &  & $E$  & ${\cal O}_{V_x}+{\cal O}_{V_y}$ & - & $w_{00}-w_{20}$ \\ \hline
\multirow{2}{*}{$C_{2v}$} & \multirow{2}{*}{(1,1,0)} & $A_1$ & ${\cal O}_{V_x}+{\cal O}_{V_y}$ & ${\cal O}_{\pi_{\rm D} \pi_{\rm D}}$ & $w_{00}-w_{20}-i\sqrt{6}w_{22}$\\
                          & & $B_1$ & ${\cal O}_{V_z}$ & - & $w_{00}+2w_{20}$ \\ \hline
\multirow{2}{*}{$C_{3v}$} & \multirow{2}{*}{(1,1,1)} & $A_1$ & ${\cal O}_{V_x}+{\cal O}_{V_y}+{\cal O}_{V_z}$ & ${\cal O}_{\pi_{\rm D} \pi_{\rm D}}$ & $w_{00}-w_{20}+i\sqrt{6}w_{22}$ \\
                          & & $E$ & $2{\cal O}_{V_x}-{\cal O}_{V_y}-{\cal O}_{V_z}$ & - & $w_{00}+i\sqrt{6}w_{22}$ \\ \hline
\end{tabular}
\label{tab:ops}
\end{table}

The general structure of $2$-point correlation functions of interest takes the form
\beq
C_{ij}(t) = \langle {\cal O}_i(t) {\cal O}^\dagger_j(0) \rangle\,,
\eeq
where the subscripts $i,\,j$ denote the operators in \Tab{ops}. We consider separately the correlation functions involving the $A_1$ and other irreducible representations. Where only one contribution exists, we measure the correlation function built from a single meson operator,  with $\Gamma_\mu =\gamma_\mu$, and extract the lowest energy eigenvalue by performing a single-exponential fit to the lattice data at large Euclidean times, $t$, which,
due to the periodicity in the temporal direction,  has the asymptotic form 
\beq
C(t) \longrightarrow c_0^2 \left(
e^{-E_0 t} + e^{-E_0 (T-t)}
\right)\,,
\eeq
where $c_0$ is the matrix element of the vacuum-to-ground states and $T$ is the total  time extent. 
The optimal fitting range is chosen so that the effective mass, $m_{\rm eff}\equiv \log \left( C(t)/C(t+1) \right)$, shows a plateau. We check that the fit yields an acceptable value for  the reduced $\chi^2$.

For the $A_1$ irreducible representation, we construct a $3\times 3$ correlation matrix. Wick contractions lead to multiple fermion-flow diagrams, involving the $\pi_{\rm D}\pi_{\rm D}$ operators, and  dependent on the assignment of momenta for $\pi_{\rm D}$.  In particular, one finds two diagrams for the cross-correlator $C_{V^{ R},\,\pi_{\rm D}\pi_{\rm D}} =  -C_{\pi_{\rm D}\pi_{\rm D},\,V^{R}}$,  which are purely imaginary and have opposite sign, and six diagrams for $C_{\pi_{\rm D}\pi_{\rm D},\,\pi_{\rm D}\pi_{\rm D}}$,  e.g. see Refs.~\cite{Alexandrou:2017mpi,CP-PACS:2007wro,Drach:2020wux}.  Once the correlation functions are measured, we solve the eigenvalue problem and extract the energy eigenvalues via a variational analysis, which assumes that the $k$-th eigenvalue is dominated by the $k$-th state with exponentially suppressed corrections from the $n$-th states with $n>k$, with $\lambda_k(t)\propto e^{-E_k t}\left(1+{\cal O}\left(e^{-E_{k+1}}\right)\right)$. 

The  phase shift  $\delta_1$ associated with the spin-1 channel is  encoded in the finite-volume energy shifts ~\cite{Luscher:1986pf,Luscher:1991cf,Rummukainen:1995vs}.  The quantisation condition yields
\beq
\det \left[
{\mathbf 1}+it_1(s)\left(
{\mathbf 1} + i {\cal M}(E,{\vec{P};L}
\right)
\right]=0,
\eeq
where the infinite-volume scattering amplitude, $t_1(s)$, is related to $\delta_1$ by $t_1(s)=i/\left(\cot\delta_1(s)-i\right)$, with $s=E^2-|\vec{P}|^2$ the invariant mass. The elements of the non-trivial matrix, ${\cal M}$, are geometric functions that map the infinite volume on the cubic lattice---\Tab{ops} \cite{Gockeler:2012yj,Alexandrou:2017mpi}. We use the short-handed notation $w_{lm}$, first introduced in Ref.~\cite{Rummukainen:1995vs}, which is a combination of the generalized Zeta function and kinematic variables. Only centre-of-mass (cm) energies in the region of elastic scattering, $2m_{\pi_{\rm D}} < E_{\rm cm} < 4m_{\pi_{\rm D}}$, enter L\"uscher's formula. 

\begin{figure*}[t]
\centering
\includegraphics[width=0.68\textwidth]{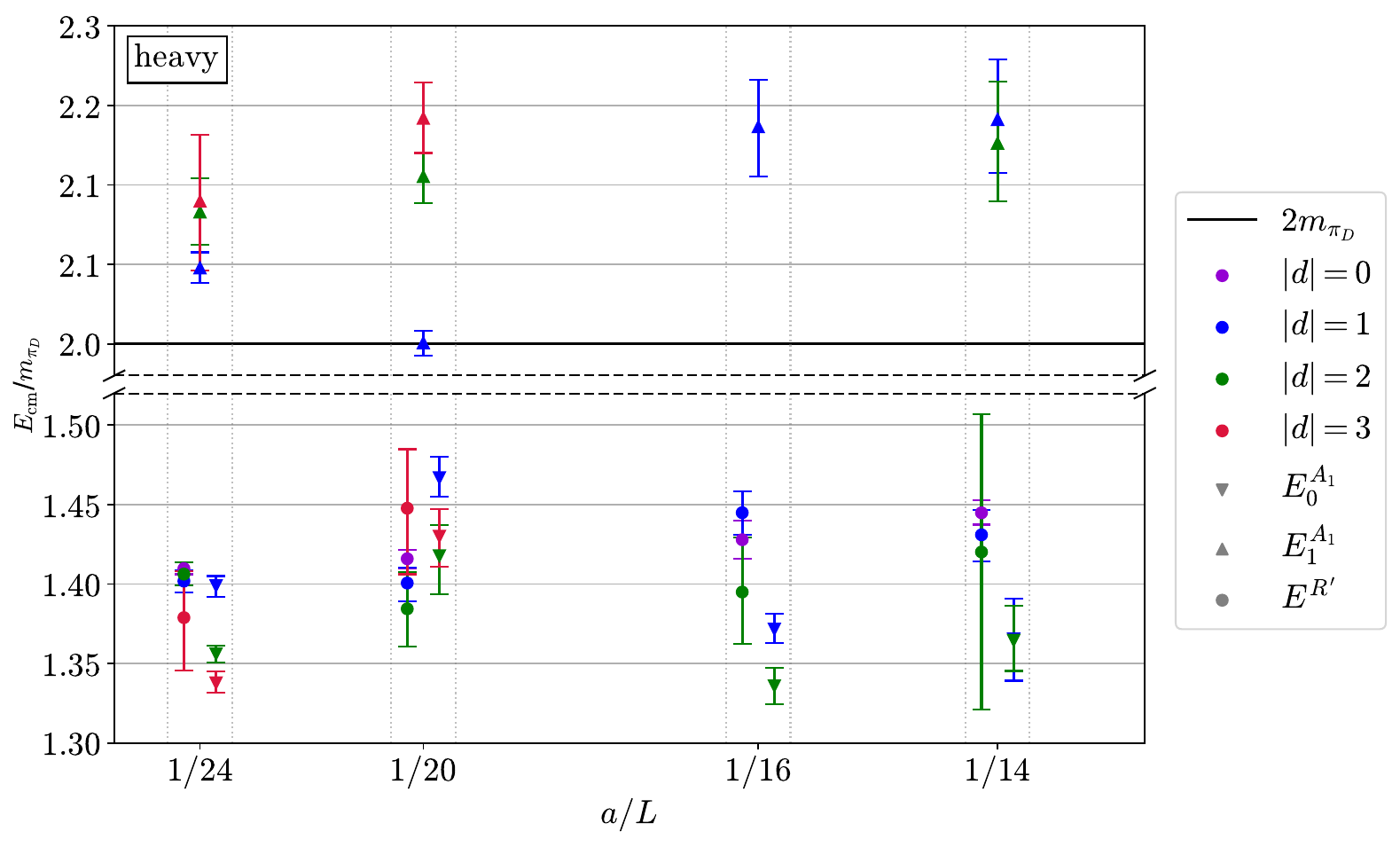}
\includegraphics[width=0.68\textwidth]{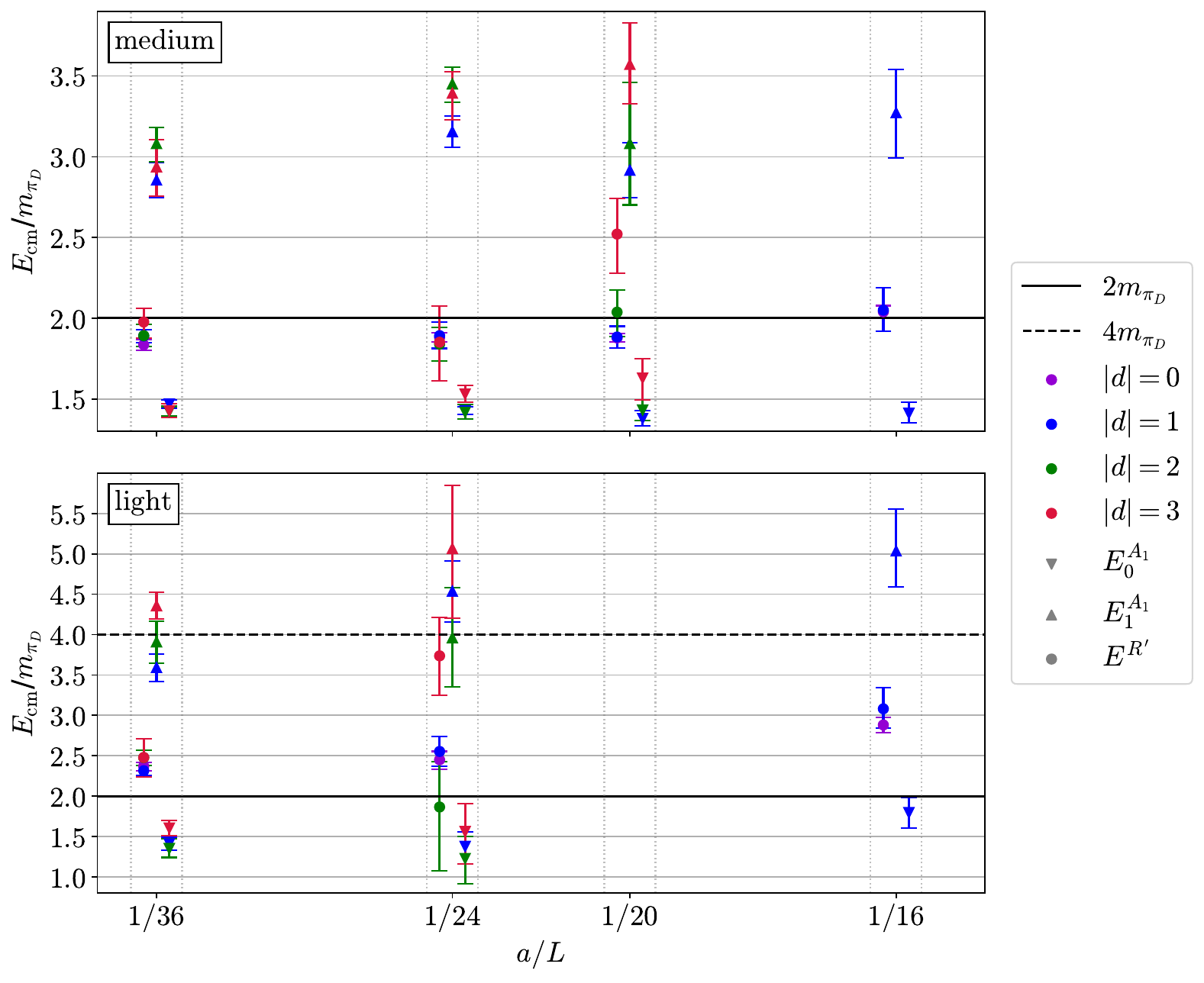}
\caption{
Finite-volume energy spectra, $E_{\rm cm}/m_{\pi_D}$, normalisaed to the pNGB masses, in the vector channel (${\bf 10}$ of $Sp(4)$) in ensembles generated with three choices of lattice parameters:  $(\beta,a\,m_0)=(6.9,0.92)$ (top), $(7.05,0.863)$ (middle), and $(7.05,0.867)$ (bottom).
Different colours are for the different assignments of the total momentum of the system in lattice units, $\vec{d}=(L/2\pi)\vec{P}$, as shown in the legends.  The filled down-pointing and up-pointing triangles denote ground state and first excited state energies in the $A_1$ channel, while the filled circles are for the ground state energies, extracted from single meson operators, in the other channels with $R'=T_1,~E,~B_1$. The solid and dashed lines represent the $2$- and $4$-pNGB thresholds, respectively.}
\label{fig:e_cm}
\end{figure*}

\section{Numerical results and discussion}

\begin{figure}[t]
\centering
\includegraphics[width=0.75\textwidth]{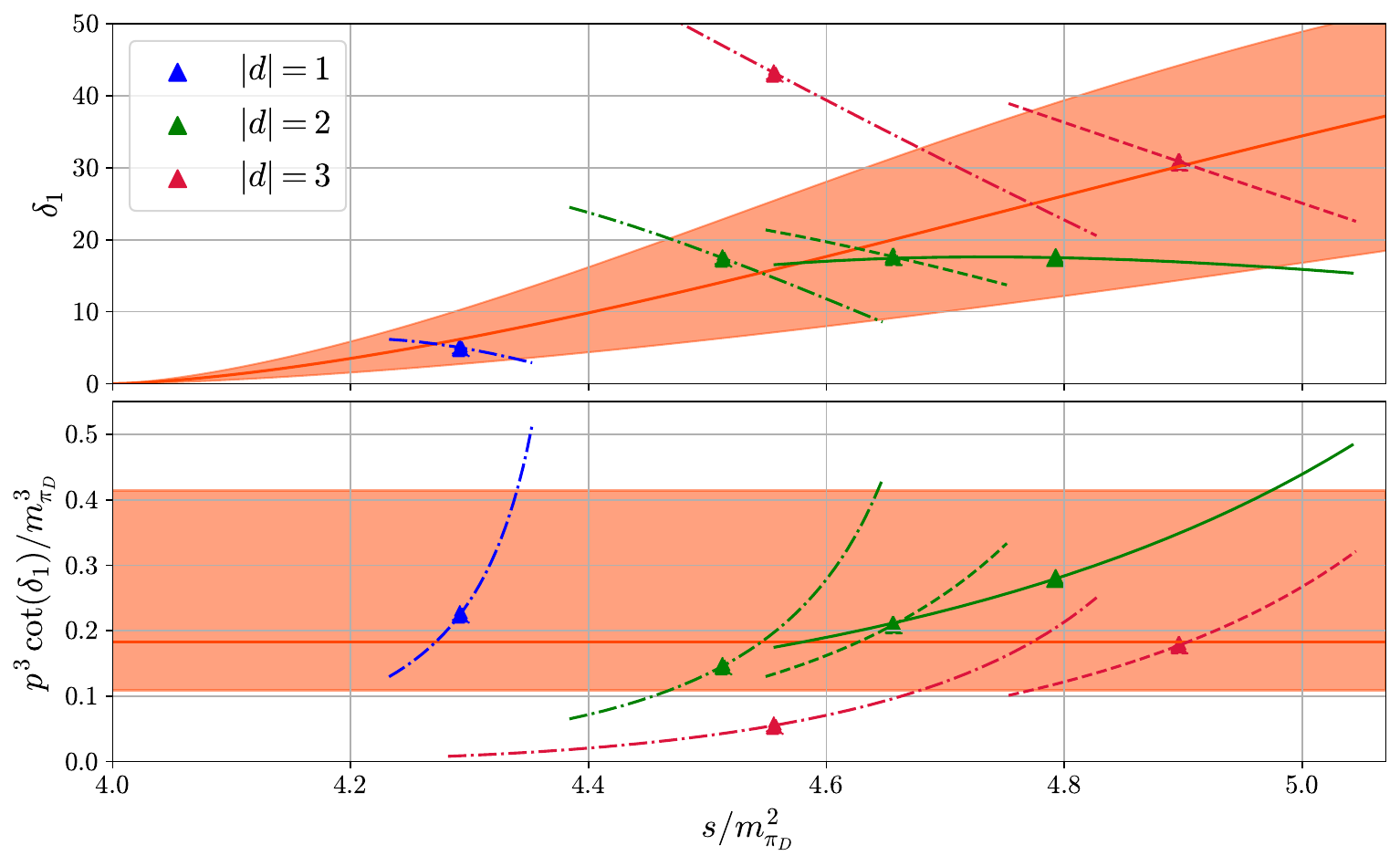}
\includegraphics[width=0.75\textwidth]{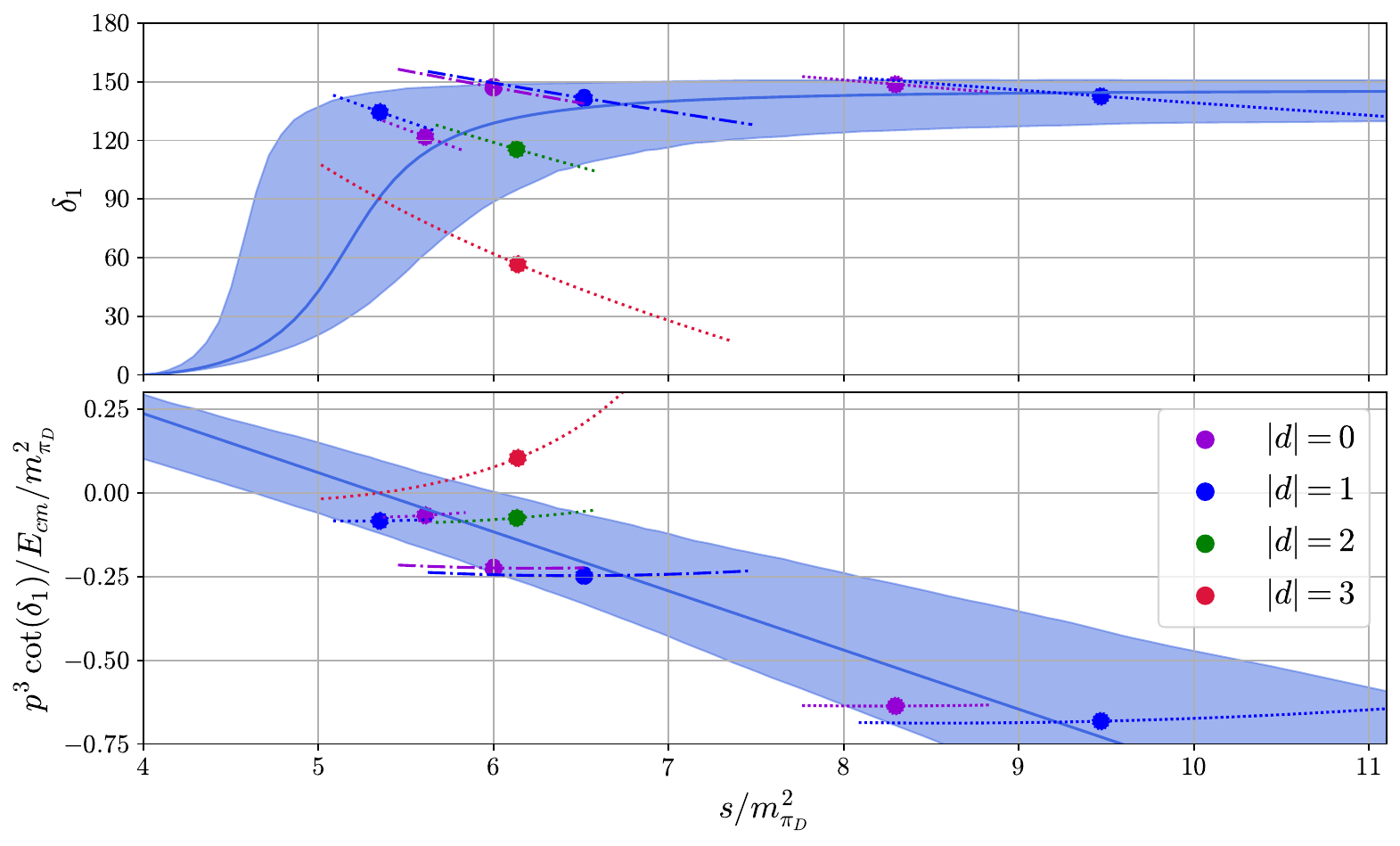}
\caption{
The infinite-volume phase shift, $\delta_1$, in the spin-1 channel for the \textit{heavy} (upper panel) and \textit{light} (lower) cases.  Different colours are for the different assignments of the total momentum of the system in lattice units, $\vec{d}=(L/2\pi)\vec{P}$, as shown in the legends. The shaded bands are the fit results obtained using the effective range expansion at the leading order and the Breit-Wigner form, respectively, in the \textit{heavy} and \textit{light} cases.}
\label{fig:delta1}
\end{figure}

For the purposes of the numerical study presented here, we discretise the action on a hypercubic lattice, with anti-periodic boundary conditions for the fermions in the time direction and periodic in all other cases, and adopt the standard Wilson plaquette action for bosons, with the Wilson-Dirac formulation of the fermions~\cite{Bennett:2017kga,Bennett:2019jzz,Bennett:2023rsl} (see also Ref.~\cite{Bennett:2023wjw} for a comprehensive review). 
We generate gauge ensembles using the hybrid Monte Carlo (HMC) algorithm, with 
Hasenbusch acceleration implemented in the HiRep code~\cite{DelDebbio:2008zf,Bussone:2018mzi,hirep}, adapted to the symplectic gauge theories~\cite{Bennett:2017kga}. The code supports the desired operators and was initially used for the $SU(2)$ gauge theory with the same matter content~\cite{Drach:2020wux}. We extended the operator basis by adding $\mathcal{O}_{V_\mu}$ with $\Gamma_\mu = \gamma_0 \gamma_\mu$ to the existing code \cite{hirep}.
We consider three distinct sets of lattice parameters, $(\beta,am_0)=(6.9,0.92),\,(7.05,0.863),\,(7.05,0.867)$, which we refer to as heavy, medium,  and light, respectively. For each, we generated ensembles for three or four different lattice volumes.  We use the Wilson flow to set the scale and monitor the topological charge \cite{Bennett:2022ftz}. The full details of the ensembles will be reported in Ref.~\cite{TELOS:2026}. For the measurement of the correlators, we employ $Z_2\times Z_2$ stochastic noise sources, and sequential sources following \cite{Alexandrou:2017mpi,CP-PACS:2007wro,Drach:2020wux} for the 2-pNGB operators. 

Our  spectral results are presented in \Fig{e_cm}, showing different lattice volumes and parameters, and displaying together different energy levels. The ground states are below the $2\pi_{\rm D}$ thresholds and thus stable in all three cases. 
In the \textit{heavy} case, the ground states have good overlap with the operators considered. 
In the other two cases, the single meson operator is associated with a significantly larger value of the energy, 
indicating that it has little overlap with the desired ground state. 

\begin{figure}[t]
\begin{center}
\centering
\includegraphics[width=0.75\textwidth]{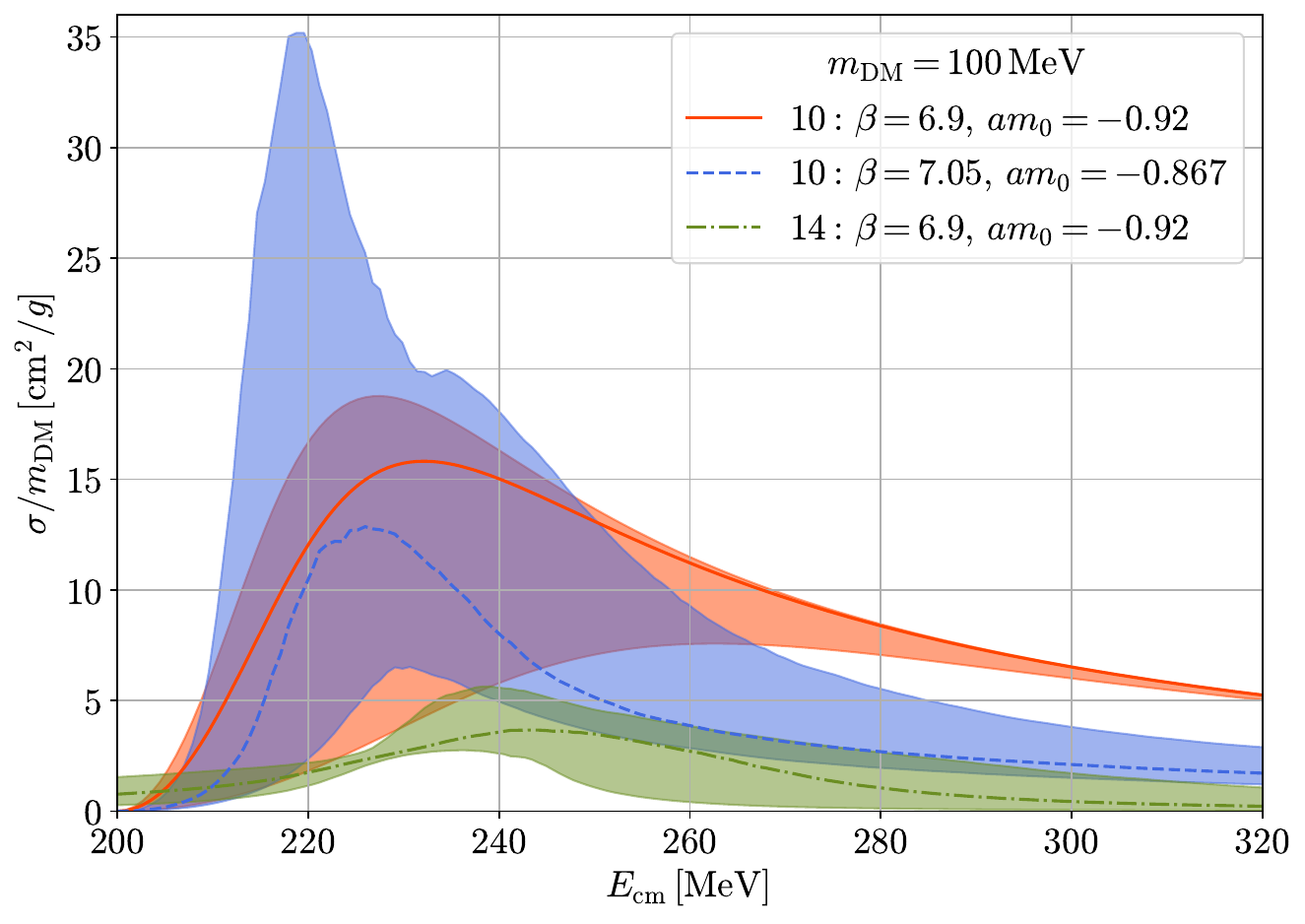}
\caption{
Cross sections for pNGB/DM scattering in the channels characterised by the ${\bf 10}$ and ${\bf 14}$ of $Sp(4)$)  channels, assuming the pNGBs act as dark matter particles with mass $m_{\rm DM}=100$ MeV.  Solid, dashed, and dot-dashed lines denote central values, while the shaded regions represent the statistical uncertainties. The result for the ${\bf 14}$ is taken from Ref.~\cite{Dengler:2024maq}.
}
\label{fig:crossx}
\end{center}
\end{figure}

In \Fig{delta1}, we show the results of the phase shift, $\delta_1$, extracted via Lüscher's finite volume analysis. In the \textit{heavy} case, the shifts of the momentum are relatively small, as shown in~\Fig{e_cm}, and so are the phase shifts. We therefore employ the effective range expansion (ERE) for spin-1 scattering to analyse the results:
\beq
p^3 \cot \delta_1 = -\frac{1}{a_1^3}+ \frac{1}{2r_1} p^2+{\cal O}\left(p^4\right)\,,
\eeq
where $p=|\vec{p}|$ is the momentum in the centre-of-mass frame with $s = 4(m_{\pi_D}^2 + p^2)$, while $a_1$ and $r_1$ are the scattering length and the effective range, respectively. 
Given statistical uncertainties, we find no meaningful contributions except for the leading term, and estimate $a_1 m_{\pi_D}=-1.76^{+0.11}_{-0.47}$, 
by performing a constant fit to the data. The fit results are shown by shaded bands in the figure. 
Note that the negative value of $a_1$ indicates that the ${\bf 10}$ channel is attractive, 
in contrast to the results for the scattering in the ${\bf 14}$ channel~\cite{Dengler:2024maq}. 
 
In the \textit{light} case, we find evidence of a resonance. Hence, we attempt to fit the data using the Breit-Wigner form
\beq
\cot \delta_1 = \frac{m_{\rho'_{\rm D}}^2-s}{\sqrt{s}\Gamma(s)}\,,
\eeq
with the width computed as a tree level process in the hidden local symmetry language~\cite{Harada:2003jx}
\beq
 \Gamma(s) = \frac{(g_{\rho'_{\rm D} {\rm \pi_D \pi_D}})^2}{6\pi}\frac{p^3}{s}\,,
\eeq
and where $g_{\rho'_{\rm D} {\rm \pi_D \pi_D}}$ is the coupling between the vector resonant state and the two pNGBs.  From the best fit, we obtain $g_{\rho'_{\rm D} {\rm \pi_D \pi_D}}=10.3^{+1.6}_{-1.0}$. 

For phenomenological considerations related to SIMP dark matter models, the scattering cross section, $\sigma$, is a most interesting quantity. Using our measurements of $\delta_1$ as inputs for the partial-wave cross section, $\sigma_\ell=(2\ell+1)4\pi\sin^2(\delta_\ell)/p^2$, we calculate $\sigma/m_{\rm DM}$ in the ${\bf 10}$ channel, taking the dark matter pNGBs to have mass $m_{\rm DM}=m_{\pi_{\rm D}}=m_{\rm PS}=100$\,MeV. 
The results are shown in \Fig{crossx}, and compared to the  ${\bf 14}$ channel, taken from Ref.~\cite{Dengler:2024maq}. The typical value of $E_{\rm cm}$ relevant to $2\rightarrow 2$ dark matter scattering in the galactic halos would be less than one MeV. 
In this kinematical region, the dominant contribution arises from the  ${\bf 14}$  channel, 
whereas the ${\bf 10}$  channel is highly suppressed.  Nevertheless, the appearance of a prominent peak at higher energies might result in significant contributions in other kinematical regions. 
To test the validity of the SIMP dark matter scenario in this model in a  realistic setting it is necessary to extend the current work in the future, both by taking into account  lattice systematics as well as by
exploring a broader region of  parameter space.

\section*{Acknowledgement}
%EB and BL are supported by the EPSRC ExCALIBUR programme ExaTEPP project EP/X017168/1. 
EB is supported by the STFC Research Software Engineering Fellowship EP/V052489/1. 
EB, BL, MP and FZ are supported by the STFC Consolidated Grant No. ST/X000648/1.
YD is supported by the Austrian Science Fund research teams grant STRONG-DM (FG1). 
DKH is supported by Basic Science Research Program through the National Research Foundation of Korea (NRF) funded by the Ministry of Education (NRF-2017R1D1A1B06033701) and by the NRF grant 2021R1A4A5031460 funded by the Korean government (MSIT). 
J-WL is supported by IBS under the project code, IBS-R018-D1. 
HH and C-JDL acknowledge support from NSTC Taiwan, through grant number 112-2112-M-A49-021-MY3. 
C-JDL is also supported by the Taiwanese MoST grant 109-2112-M-009-006-MY3. 
%C-JDL also receives support from two other NSTC grants, 112-2639-M-002-006-ASP and 113-2119-M-007-013.
BL and MP have been supported by the STFC  Consolidated Grant No. ST/T000813/1 and received funding from the European Research Council (ERC) under the European Union’s Horizon 2020 research and innovation program under Grant Agreement No.~813942. 
BL was supported by the STFC Consolidated Grant No. ST/X00063X/1. 
DV is supported by STFC under Consolidated Grant No. ST/X000680/1.

{\bf High performance computing}---This work used the DiRAC Data Intensive service (CSD3) at the University of Cambridge,  the DiRAC Data Intensive service (DIaL3) at the University of Leicester and the DiRAC Extreme Scaling service (Tursa) at the University of Edinburgh, managed respectively by the University of Cambridge University Information Services, the University of Leicester Research Computing Service and by EPCC on behalf of the STFC DiRAC HPC Facility (www.dirac.ac.uk). The DiRAC service at Cambridge, Leicester, and Edinburgh are funded by BEIS, UKRI and STFC capital funding and STFC operations grants. DiRAC is part of the UKRI Digital Research Infrastructure.

Numerical simulations have been performed on the Swansea SUNBIRD cluster (part of the Supercomputing Wales project) and on the NURION at KISTI. The Swansea SUNBIRD system is part funded by the European Regional Development Fund (ERDF) via Welsh Government. The NURION system is  supported by the National Supercomputing Center with supercomputing resources including technical support (KSC-2023-CRE-0549 and KSC-2024-CRE-0530).
Calculations were performed using supercomputer resources provided by the Austrian Scientific Computing center (ASC), in  particular using the Vienna Scientific Cluster (VSC4).

%{\bf Research Data Access Statement}---The data generated and the analysis code for this manuscript can be downloaded from  Ref.~\cite{workflow_release,data_release}. 

%{\bf Open Access Statement}---For the purpose of open access, the authors have applied a Creative Commons 
%Attribution (CC BY) license to any Author Accepted Manuscript version arising.

\end{document}